# Generation and Confirmation of a (100x100)-dimensional entangled Quantum System

Mario Krenn[1,2,*], Marcus Huber[3,4,5], Robert Fickler[1,2], Radek Lapkiewicz[1,2], Sven Ramelow[1,2,6], Anton Zeilinger[1,2,*]

**Affiliations:**

[1]Institute for Quantum Optics and Quantum Information (IQOQI), Austrian Academy of Sciences, Boltzmanngasse 3, A-1090 Vienna, Austria.

[2]Vienna Center for Quantum Science and Technology (VCQ), Faculty of Physics, University of Vienna, Boltzmanngasse 5, A-1090 Vienna, Austria.

[3]University of Bristol, Department of Mathematics, Bristol BS8 1TW, U.K.

[4]ICFO-Institut de Ciencies Fotoniques, E-08860 Castelldefels (Barcelona), Spain.

[5]Fisica Teorica: Informacio i Fenomens Quantics, Departament de Fisica, Universitat Autonoma de Barcelona, E-08193 Bellaterra (Barcelona), Spain.

[6]Cornell University, 271 Clark Hall, 142 Science Dr., Ithaca, 14853 NY, USA (current address).

*Correspondence to: M.K. (mario.krenn@univie.ac.at) and A.Z. (anton.zeilinger@univie.ac.at)

**Abstract**: Entangled quantum systems have properties that have fundamentally overthrown the classical worldview. Increasing the complexity of entangled states by expanding their dimensionality allows the implementation of novel fundamental tests of nature, and moreover also enables genuinely new protocols for quantum information processing. Here we present the creation of a (100x100)-dimensional entangled quantum system, using spatial modes of photons. For its verification we develop a novel nonlinear criterion which infers entanglement dimensionality of a global state by using only information about its subspace correlations. This allows very practical experimental implementation as well as highly efficient extraction of entanglement dimensionality information. Applications in quantum cryptography and other protocols are very promising.

Quantum entanglement of distant particles leads to correlations that cannot be explained in a local realistic way (1-3). To obtain a deeper understanding of entanglement itself, as well as its application in various quantum information tasks, increasing the complexity of entangled systems is important. Essentially, this can be done in two ways: The first method is to increase the number of particles involved in the entanglement (4). The alternative method is to increase the entanglement dimensionality of a system.

Here we focus on the latter one, namely on the dimension of the entanglement. The text is structured as follows: After a short review of properties and previous experiments, we present a new method to verify high-dimensional entanglement. Then we show how we experimentally create our high-dimensional 2-photon entangled state. We analyze this state with our new method and verify a (100x100)-dimensional entangled quantum system. We conclude with a short outlook to potential future investigation.

High-dimensional entanglement provides a higher information density than conventional two-dimensional (qubit) entangled states, which has important advantages in quantum communication: Firstly, it can be used to increase the channel-capacity via superdense-coding (5). Secondly, high-dimensional entanglement enables the implementation of quantum communication tasks in regimes where mere qubit entanglement does not suffice. This involves situations with a high level of noise from the environment (6,7), or quantum cryptographic systems where an eavesdropper has manipulated the random number generator involved (8). Moreover, the entangled dimensions of the whole Hilbert space also play a very interesting role in quantum computation: High-dimensional systems can be used to simplify the implementation of quantum logic (9). Furthermore, it has been found recently (10) that any continuous measure of entanglement (such as Concurrence, Entanglement of Formation or Negativity) can be very small, while the quantum system still permits an exponential computation speedup over classical machines. This is not the case for the dimension of entanglement – for every quantum computation it needs to be high (11,12), which is another hint at the fundamental relevance of the concept.

So far, high-dimensional entanglement has been implemented only in photonic systems. There, different multi-level degrees of freedom such as spatial modes (13), time-energy (14), path (15) (16) as well as continuous variables (17) (18) have been employed. Especially entanglement of spatial modes of photons has attracted much attention in recent years (19) (20) (21) (22) (23) (24) (25) (26) (27) (28), because it is readily available from optical nonlinear crystals and the number of involved modes of the entanglement can be very high (29).

In a recent experiment the non-separability of a two-photon state was shown, by observing EPR correlations of photon pairs in down-conversion (30) (for a similar experiment, see (31)). The authors were able to observe entanglement of ~2500 spatial states with a camera. In our experiment we go a step further and not only show non-separability, but we can also extract information about the dimensionality of the entanglement. Precisely, we experimentally verify 100-dimensional entanglement.

One main challenge that remains is the detection and verification of high-dimensional entanglement. For reconstructing the full quantum state via state tomography, the number of required measurements is impractical even for relatively low dimensions because it scales quadratically with the quantum system dimension (24) (27). Even if one had reconstructed the full quantum state, the quantification of the entangled dimensions is a daunting task, analytically and even numerically (32). If the full density matrix of the state is not known, it is only possible to give lower bounds of the entangled dimensions. Such methods are usually referred to as a Schmidt number witness (33) (34) (35).

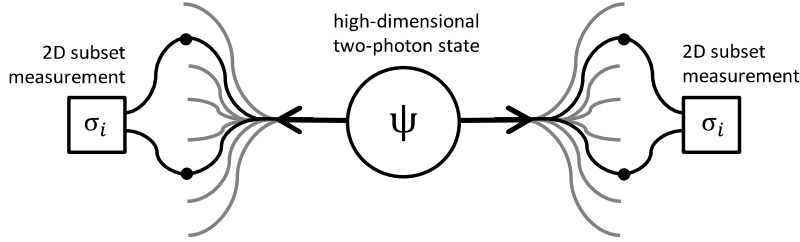

**Figure 1:** Visualization of the measurement concept. The two photons are sent in two different directions. Each of the photons is in a mixture of many modes. We perform the same two-dimensional subspace measurement on both photons. When we consider all two-dimensional subspaces, we can determine the dimensionality of entanglement.

Strictly speaking, we found a mathematically well-defined and intuitively reasonable method that answers the following question: *For a given high-dimensional two-photon state, if correlations between D dimensions of each photon are measured, what is the minimum necessary entanglement dimensionality d required to explain the correlations?* (The precise mathematical formulation of this question is given in the supplementary.)

Our approach works such that we define a measurable witness-like quantity W and search for the d-dimensional entangled state maximizing it. When we perform the measurement and exceed the maximal value, we know that the measured quantum state was at least (d+1)-dimensional entangled. This approach is a generalization of conventional entanglement witnesses, which define a boundary between separable and entangled states. We not only want a boundary between separable and entangled states, but also between different dimensions of entanglement.

The main idea is to look at 2-dimensional subspaces, and therewith measure the correlations of the two photons. In analogy to two-dimensional systems (such as photon's polarization), we can measure the visibilities in three mutually unbiased bases (MUBs). Mathematically, the visibility is defined as $V_i = |\langle \sigma_i \otimes \sigma_i \rangle|$, i={x,y,z}, where $\sigma_i$ denote the single-qubit Pauli matrices (for polarization $\sigma_x$, $\sigma_y$ and $\sigma_z$ represent measurement in the diagonal/antidiagonal, left/right and horizontal/vertical basis). The concept of the measurement is illustrated in Figure 1. With these measurements, it is possible to detect entanglement between two-dimensional subsystems (35). What we found is a way how such measurements in all two-dimensional subsystems imply a lower bound of the entangled dimensions in the whole quantum system.

In our experiment we are in a regime where it is unfeasible to reconstruct the full density matrix because of the required number of measurements due to the high dimension. Therefore we can only identify lower bounds of the entangled dimensions. Furthermore, all previously published methods for extracting the dimension of entanglement turned out to be impractical for our system. They usually require access to observables on the full Hilbert space, which were not available for our experiment. For these reasons we were required to develop a novel approach.

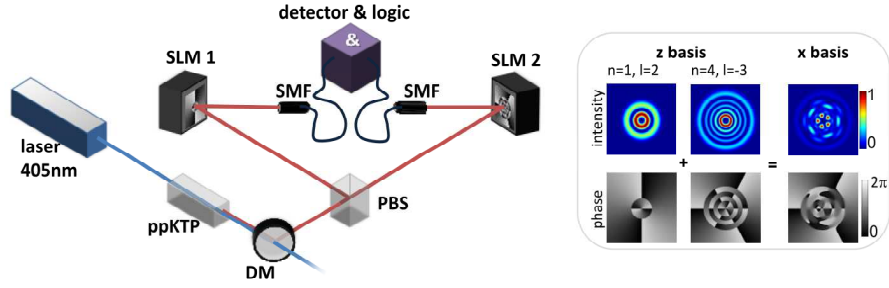

**Figure 2:** Schematic of the experimental setup. We pump a type-II nonlinear periodically poled potassium titanyl phosphate (ppKTP) crystal with a 405 nm, ~40mW single-mode laser. Spontaneous parametric down-conversion creates collinear photon pairs with 810 nm wavelength and orthogonal polarizations. We remove the pump beam at a dichroic mirror (DM) and separate the two photons at a polarizing beam splitter (PBS). In both arms of the setup we use spatial light modulators (SLM) to perform a mode-transformation of the photons. The transformation done by a computer-generated hologram at the SLM converts a specific mode into the fundamental Gauss mode. Only the Gauss mode couples into a single mode fiber (SMF), thus the SLM+SMF combination acts as a mode filter (39). In the end, we detect the photons with avalanche photo-diode based single photon detectors and analyze the time correlation using a coincidence-logic. The inset shows an example of a two-dimensional subspace. The intensities and phases for two different modes in the z-basis are demonstrated, and their superposition leads to a mode in the x-basis. The y-basis can be constructed similarly.

Our measureable quantity W is the sum of the visibilities in all 2-dimensional subspaces

$$W = \sum_{a=0}^{D-2} \sum_{b=a+1}^{D-1} \frac{\left(V_x^{a,b} + V_y^{a,b} + V_z^{a,b}\right)}{N_{a,b}}, \qquad (1)$$

where a and b stand for specific states of the photons, D is defined as above – it stands for the number of modes considered. $V_i^{a,b}$ stand for the visibility in basis *i*. The $N_{a,b}$ stands for the normalization. It is the source of the non-linearity of W which leads to convenient experimental properties, however makes it very difficult in general to handle mathematically. That nonlinearity is responsible for the fact that the measurement results are automatically normalized (i.e. all visibilities can go up to one), because by measuring in 2-dimensional subspaces, we ignore all other modes (see supplementary). Therefore we do not need to renormalize our measurement results in any way afterwards. Non-linear entanglement witnesses have already been used in earlier experiments and demonstrated specific advantages over linear witnesses (36) (37).

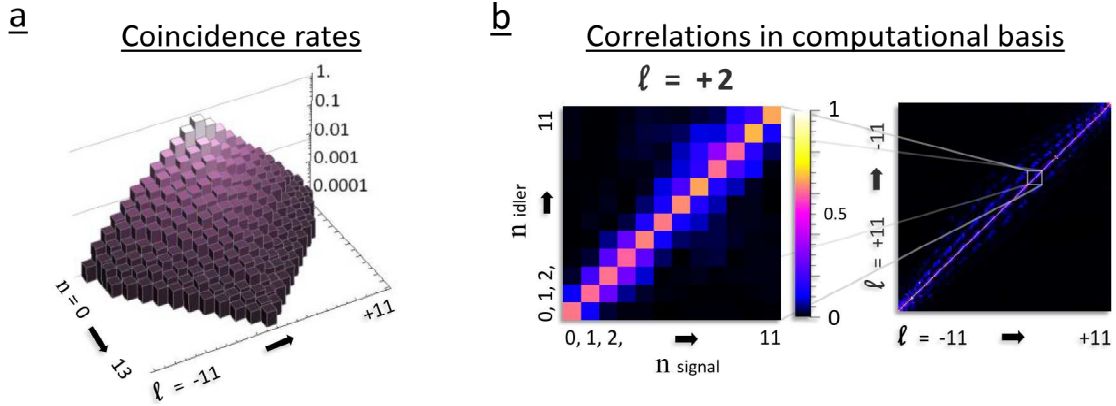

**Figure 3: a)** Normalized coincidence rate of different modes (with logarithmic scale), depending on the two mode numbers ("full-field"-bandwidth). The absolute count rate was 105.500 photon pairs per second for a pump power of 60mW. To be precise, this is the summed count rate of all 186 modes, not taking into account the inefficiencies of the detectors or imperfect coupling into single-mode fibres. **b)** Weighted correlations between different modes in z-basis. Due to different probabilities of different modes, in these pictures we weight every correlation with the probability of the modes involved. That means, $\langle i|j\rangle_{weighted} = N \frac{\langle i|j\rangle_{measured}}{\sqrt{\langle i|i\rangle\langle j|j\rangle}}$, where $i$ and $j$ stand for different two-photon modes, and N is a normalization constant. The left part shows the correlation of modes with l=2, and reveals good correlation of modes with the same number of radial nodes. The right picture visualizes all correlations in the z-basis.

Next we search for the d-dimensional entangled state which is maximizing the quantity W in eq. 1. The maximization was not yet possible in general (which stays an interesting open problem, especially for more realistic experimental situations). However we maximized W for a very big and particularly important class which we believe to be sufficient for our experiment. In other words, we used a physical assumption about our state in the derivation, which we will explain in more details later. The basis of the maximization is a combination of the method of Lagrange multipliers and algebraic considerations. This enables us to find the maximizing d-dimensional quantum state for the quantity W (see supplementary material), and implies an upper bound on the quantity in eq. 1 for d-dimensional entangled states, which can be written as

$$W \leq 3\frac{D(D-1)}{2} - D(D-d), \qquad (2)$$

If the measurements exceed the bound, the quantum state was at least (d+1)-dimensional entangled. Otherwise, if the inequality is fulfilled (W is smaller than the right side), we cannot make a statement about the dimensionality of entanglement.

The bounds can be understood intuitively. A maximally entangled state in D dimensions will have a visibility of one in all three MUBs in every two-dimensional subspace. This is represented by the first term on the right side. If the entanglement dimensionality of the state is smaller than that of the observed Hilbert-space, the maximally reachable value decreases by D for each non-entangled dimension (D-d) – which is expressed by the second term.

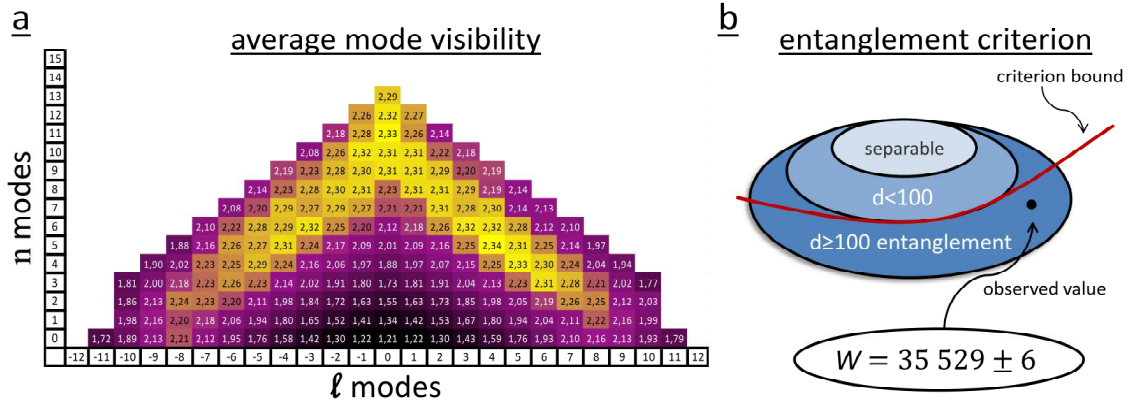

**Figure 4: a)** The average sum of visibilities (in x, y, and z basis) of a specific mode with all other modes is shown. The observable structure originates from non-maximally entanglement due to different count rates for different modes (Fig. 3a). The bright regions in the center are modes with a similar probability. The central low-order modes (such as the Gauss mode) have the highest probability, therefore the lowest average visibility. Precisely, the Gauss mode has an average sum of visibilities of 1,21 – which mainly results from the visibilities in z-basis. A maximally entangled high-dimensional state would have a summed visibility of 3 for every mode pair. The structure shows that different modes contribute differently to the entanglement criterion. **b)** To reveal information about the global, high-dimensional entanglement, we have to calculate the value for the quantity W in eq. (1). This is done by summing up all visibilities of all subsets. Here, the concept and result of the entanglement dimensionality criterion is visualized. With our criterion in eq. (2), quantum states are divided into two parts – those with entanglement dimensionality smaller than 100 and those with a larger one (red line). We have observed a value that lies in the lower part, thus we have verified 100-dimensional entanglement.

The quantity in eq. 1 is remarkable because the number of required measurements scales only linearly with the dimension of the whole Hilbert space, in contrast to state tomography, which scales quadratically. Furthermore, it only involves measurements in 2-dimensional subspaces, which are easier to implement than general high-dimensional measurements. Moreover, the quantity W in eq. 1 is non-linear, which makes it particularly efficient for non-maximally entangled quantum states (see supplementary).

In our experiment, we apply this novel method to a two-photon quantum system. The photon pair is created by pumping a nonlinear crystal with a laser, where spontaneous parametric down-conversion (SPDC) occurs. For the high-dimensional degree of freedom we use spatial modes of light. Specifically, we use the Laguerre-Gauss (LG) basis to analyse entanglement. LG modes form a basis of solutions of the paraxial wave equation in the cylindrical coordinate system. They are described by two quantum numbers. One quantum number $l$ corresponds to the orbital angular momentum momentum (OAM, or equivalently, the topological charge) of the photon (38) (39). The second quantum number $n$ corresponds to the radial nodes in the intensity profile. Only lately this second degree of freedom has been analyzed theoretically in a quantum mechanical framework (40) (41) (42).

In the down-conversion process the angular momentum of the photons is conserved, therefore this degree of freedom is anti-correlated. For the radial quantum number $n$ the situation is more complicated. The full down-conversion process concerning the correlations for the radial quantum number has been analyzed in detail (40) and quasi-perfect correlations have been found for specific situations. Recently, these quasi-perfect correlations have been demonstrated experimentally (43). The state we expect from down conversion can then be written as a perfectly (anti-)correlated pure state $|\psi\rangle = \sum_{n=0}^{\infty}\sum_{l=-\infty}^{\infty} a_{n,l}|LG_{n,l}, LG_{n,-l}\rangle$ with $l$ and $n$ dependent coefficients $a$. In the derivation of the bounds in eq. 2 we restricted the states to be perfectly correlated. This means we assume a physical property of our input state – namely perfect (anti-)correlation of the modes. Small deviations from this assumption (which we have observed in the experiment, see Fig 3b) have been analyzed numerically (see supplementary), and we found that it only reduces our observed W, thus justifies the application of the criterion in eq. 2 in our experiment. The full analytical treatment in the general case is an interesting open problem.

The experimental analysis of the LG modes of the photon pair produced is done by a holographic mode transformation using a spatial light modulator. With that we can transform any desired mode to a Gauss mode. By using a single mode fiber, we filter only for Gauss modes and thereby project the quantum state into the desired mode (39). The setup and exemplary LG modes are shown in Fig. 2.

In our experiment we analyze the correlations of 186 modes of two photons. The number of modes, 186 in our case, corresponds to D in eq. 2. We employ LG modes with an angular quantum number up to $l$=11, and a radial quantum number up to $n$=13. To calculate the quantity in eq. 1, we need to measure in every 2-dimensional subspace (there are 186x185/2 = 17205 2-dimension subspaces) the visibility in x-, y- and z-basis, which corresponds to 3x4 measurement per subspace. Altogether this results in ~200 000 measurements (with ~750 million detected photon pairs). For comparison, if we had performed a full state tomography, we would have needed to perform more than 1 billion measurements. When we sum up all of our measured visibilities according to eq. 1, we find

$$W_{D=186} = 35\ 529 \pm 6, \tag{3}$$

which corresponds to at least 100-dimensional entanglement according to inequality (2) (101dim: W>35619; 100dim: W>35433; 99dim: W>35247). The confidence interval corresponds to one standard deviation due to the statistical uncertainty. It has been calculated using Monte Carlo simulation assuming Poisson distribution of the count rates. The detailed measurement results and the calculation of (3) can be seen in Fig. (4) and in the Supplementary Information. The quantity W in (1) corresponds to measurements of all 2-dimensional subspaces in a DxD-dimensional quantum state. It can be seen in Fig. (4a) that some modes contribute more to the quantity than others, thus we can try to find a smaller optimal set of modes that shows the higher-dimensional entanglement. We find that by removing 19 modes (that means, not taking into account all 2-dimensional subspace measurement with them), we can find at least 103-dimensional entanglement.

One way to bring this in relation with other photonic and multipartite entanglement experiments is the following: The dimension of the entangled Hilbert-space scales with dim=$d^N$, where d stands for the entangled dimensions and N is the number of involved parties. Our experiment shows an entangled Hilbert-space dimension of dim=(103x103) ~ $2^{13.4}$ that is larger than the biggest entangled photonic Hilbert-space reported so far (with dim=$2^{10}$) (44). Interestingly, it is of similar magnitude as that of the largest quantum systems with multipartite entanglement measured so far, such as 14-qubit ion entanglement with dim=$2^{14}$ (45).

Our results show that we can experimentally access a quantum state of two photons which is at least (100x100)-dimensionally entangled. This was possible by developing a novel method to analyze efficiently and in an experimentally practical way quantum states with very high dimensions. Furthermore, we exploited the full potential of transverse spatial modes, namely both the radial and the angular quantum numbers.

Such high-dimensional entanglement offers a great potential for quantum information applications. There are situations where 2-dimensional entanglement is not sufficient anymore but high-dimensional entangled systems are able to perform the task. In realistic quantum cryptography schemes, for example where noisy environment or manipulated random number generators lead to a breakdown of the system for low-dimensional entangled states, high-dimensionality of the entanglement sustains the security (6-8). Experimental setups as presented here are suitable for such tasks. Additionally, for quantum computation it is necessary to utilize a large entangled Hilbert space for any quantum speedup. As our result shows that very high-dimensional entangled Hilbert spaces are experimentally accessible, we envision that it will trigger future experiments to solve the next important open question: How to implement experimentally arbitrary controlled transformations between spatial modes in order to realize quantum computational or similar tasks.

**Acknowledgement**
This work was supported by the European Research Council (ERC Advanced Grant No. 227844 "QIT4QAD"; SIQS, No. 600645 EU-FP7-ICT); and the Austrian Science Fund FWF with the SFB F40 (FoQus) and W1210-2 (CoQus). We would like to thank Christoph Schäff, Mehul Malik and William Plick for helpful discussions. MH would like to acknowledge the Marie Curie IEF grant QuaCoCoS – 302021. SR would like to acknowledge the EU Marie-Curie Fellowship (PIOF-GA-2012-329851). MK would like to thank the quantum information theory groups at ICFO and UAB (Barcelona), and MH for nice hospitality.

# SUPPLEMENTAL MATERIAL
# Generation and Confirmation of a (100x100)-dimensional entangled Quantum System

Mario Krenn[1,2,*], Marcus Huber[3,4,5], Robert Fickler[1,2], Radek Lapkiewicz[1,2], Sven Ramelow[1,2,6], Anton Zeilinger[1,2,*]

[1] Institute for Quantum Optics and Quantum Information (IQOQI),
Austrian Academy of Sciences, Boltzmanngasse 3, A-1090 Vienna, Austria.
[2] Vienna Center for Quantum Science and Technology (VCQ), Faculty of Physics,
University of Vienna, Boltzmanngasse 5, A-1090 Vienna, Austria.
[3] University of Bristol, Department of Mathematics, Bristol BS8 1TW, U.K.
[4] ICFO-Institut de Ciencies Fotoniques, E-08860 Castelldefels (Barcelona), Spain.
[5] Fisica Teorica: Informacio i Fenomens Quantics, Departament de Fisica,
Universitat Autonoma de Barcelona, E-08193 Bellaterra (Barcelona), Spain.
[6] Cornell University, 271 Clark Hall, 142 Science Dr., Ithaca, 14853 NY, USA (current address)
[*] Correspondence on the supplemental material to M.K. (mario.krenn@univie.ac.at) and A.Z. (anton.zeilinger@univie.ac.at)

## LAGUERRE-GAUSS MODES

Laguerre-Gauss (LG) modes are solutions of the paraxial wave equation (which, for example, describes the propagation of laser beams). They are described by two mode numbers, the radial mode number $n$, and the angular mode number $l$. The LG modes are orthogonal to each other, and form a complete basis. They can be written as

$$LG_{n,l}(r,\phi,z=0) = \frac{N_{n,l}}{w_0}\left(\frac{r\sqrt{2}}{w_0}\right)^{|l|}\exp\left(-\frac{r^2}{w_0^2}\right)L_n^{|l|}\left(\frac{2r^2}{w_0^2}\right)\exp\left(il\phi\right), \tag{S1}$$

where $N_{n,l}$ is a normalisation constant, $w_0$ is the beam waist at z=0, and $L_n^{|l|}$ are Laguerre polynomials. The $l$ number corresponds to the orbital angular momentum of the mode, and $n$ stands for the number of nodes in the intensity profile in radial direction.

## MATHEMATICAL FORMULATION OF OUR RESULT

With d-dimensional entanglement we understand correlations that can never be explained by convex combinations of pure states with a Schmidt rank lower than d. Notice that we indeed have entanglement in a 34.596 (186x186) dimensional Hilbert space between almost all possible involved dimensions. However, even if all levels are entangled, it might still be that a convex combination of qubit states would still be enough to explain these correlations. This is what we explicitly exclude in deriving the nonlinear witness for the Schmidt number, where an entanglement dimensionality of 103 signifies that no combination of d=102 states could explain the observed correlations, thus proving genuine high dimensional entanglement.

## BOUNDING THE SCHMIDT NUMBER FROM NORMALIZED SUBSPACE CORRELATIONS

The goal of this section is to the determine lower bounds on the Schmidt-number from the sum of all two-dimensional normalized subspace correlations in a $D$ dimensional system. It is organized as follows:

- We first introduce the general correlation criterion for normalized subspaces
- We continue by deriving tight lower bounds for the same correlations in non-normalized subspaces
- Finally we derive the global maximum of this correlation function for perfectly correlated modes.

Using the following abbreviation

$$|LG_{n,l}\rangle = |k\rangle , \tag{S2}$$

where we count all *LG* modes of $\{n, l\}$ (i.e. all Laguerre-Gauss modes with mode number n (radial modes) and l (orbital angular momentum modes)), we can represent the performed measurements via the following operators

$$\sigma_x^{kl} := |k\rangle\langle l| + |l\rangle\langle k|,$$
$$\sigma_y^{kl} := i|k\rangle\langle l| - i|l\rangle\langle k|,$$
$$\sigma_z^{kl} := |k\rangle\langle k| - |l\rangle\langle l|. \tag{S3}$$

In order to lower bound the dimensionality of entanglement in normalized subspaces we use the following correlation function

$$C(\rho) = \sum_{k<l} \sum_{l=1}^{D-1} g(\rho^{kl}), \tag{S4}$$

where the sum is taken over all $\rho^{kl}$, the normalized subspace density matrices, where all but two degrees of freedom on both sides are ignored, i.e.

$$\rho^{kl} := \frac{(|k\rangle\langle k| + |l\rangle\langle l|) \otimes (|k\rangle\langle k| + |l\rangle\langle l|)\rho(|k\rangle\langle k| + |l\rangle\langle l|) \otimes (|k\rangle\langle k| + |l\rangle\langle l|)}{N_{kl}}, \tag{S5}$$

where $N_{kl}$ is the normalization, such that $\mathrm{Tr}(\rho^{kl}) = 1$, and

$$g(\rho^{kl}) = \mathrm{Tr}\left((\sigma_z^{kl} \otimes \sigma_z^{kl} - \sigma_y^{kl} \otimes \sigma_y^{kl} + \sigma_x^{kl} \otimes \sigma_x^{kl})\rho^{kl}\right). \tag{S6}$$

Comparing these to the correlations on the total state, i.e.

$$f_{kl}(\rho) = \mathrm{Tr}\left((\sigma_z^{kl} \otimes \sigma_z^{kl} - \sigma_y^{kl} \otimes \sigma_y^{kl} + \sigma_x^{kl} \otimes \sigma_x^{kl})\rho\right), \tag{S7}$$

we can write

$$g(\rho_{kl}) = \frac{f_{kl}}{N_{kl}}, \tag{S8}$$

and thus

$$C(\rho) = \sum_{k<l} \sum_{l=1}^{D-1} \frac{f_{kl}}{N_{kl}}. \tag{S9}$$

It is important to note that of course we only consider contributions from subspaces with a nonzero contribution, i.e. if $N_{kl} = 0$ we set $g_{kl} = 0$.

### Lower bounds from non-normalized subspaces

The first step of the witness construction is the determination of the maximal value of $f(\rho) := \sum_{k<l} \sum_{l=1}^{D-1} f_{kl}(\rho)$ for *d*-dimensional states.

$$\sum_{k<l} \sum_{l=1}^{D-1} f_{kl} \leq \max_{|\psi_d\rangle} f(|\psi_d\rangle\langle\psi_d|), \tag{S10}$$

where $|\psi_d\rangle = \sum_{i=0}^{d-1} \lambda_i |i\rangle_A |i\rangle_B$. The first step to achieve this maximization is to realize that

$$f(|\psi\rangle\langle\psi|) \leq \mathrm{Tr}[(2D|\phi_D\rangle\langle\phi_D| + (D-3) \sum_{i=0}^{D-1} |ii\rangle\langle ii|)|\psi\rangle\langle\psi|], \tag{S11}$$

where $|\phi_D\rangle := \frac{1}{\sqrt{D}} \sum_{i=0}^{D-1} |ii\rangle$ and we got an upper bound via setting all negative contributions to 0. It can be derived [1][2] that the maximal overlap of a *D*-dimensional state $|\phi_D\rangle$ with a *d*-dimensional state $\mathrm{Tr}(|\psi_d\rangle\langle\psi_d||\phi_D\rangle\langle\phi_D|)$ is achieved by $|\psi_d\rangle := \frac{1}{\sqrt{d}} \sum_{i=0}^{d-1} |ii\rangle$. This state also maximizes $\mathrm{Tr}[((D-3) \sum_{i=0}^{D-1} |ii\rangle\langle ii|)|\psi\rangle\langle\psi|]$ such that we can infer that indeed this state achieves the global maximum for $\max_{|\psi_d\rangle} f(|\psi_d\rangle\langle\psi_d|)$ and thus we can directly calculate

$$f(\rho) \leq (2d + (D-3)). \tag{S12}$$

Due to the linearity of the $f(\rho)$, the maximum is a pure state, thus (S12) is an upper limit for pure and mixed states.

## The structure of the global maximum

Before we proceed to bound the maximum of $C(\rho)$ as a function of $D$ and $d$ we require a few observations:

$$g_{kl} = \frac{4\Re e[\langle kk|\rho|ll\rangle] + \langle kk|\rho|kk\rangle + \langle ll|\rho|ll\rangle - \langle kl|\rho|kl\rangle - \langle lk|\rho|lk\rangle}{\langle kk|\rho|kk\rangle + \langle ll|\rho|ll\rangle + \langle kl|\rho|kl\rangle + \langle lk|\rho|lk\rangle} \leq \frac{4\Re e[\langle kk|\rho|ll\rangle] + \langle kk|\rho|kk\rangle + \langle ll|\rho|ll\rangle}{\langle kk|\rho|kk\rangle + \langle ll|\rho|ll\rangle}, \quad \text{(S13)}$$

i.e. monotonically decreasing in the elements $\langle kl|\rho|kl\rangle + \langle lk|\rho|lk\rangle$ for all $k$ and $l$. We now argue that for our physical system it is sufficient to maximize over states that are perfectly correlated, i.e. states that can be written in the form of $\rho = \sum_{k,l} c_{kl}|kk\rangle\langle ll|$ with $\text{Tr}(\rho) \leq 1$. This is physically motivated and discussed in the next section. This particular form of the maximizing density matrices implies that all Schmidt decompositions can be made in computational basis and we can write every vector in the decomposition of the maximizing $\rho$ as $|\psi_\alpha\rangle = \sum_{k\in\alpha} \lambda_k^\alpha |kk\rangle$.

This further implies we can decompose the maximizing density matrix as

$$\rho_{max} = \sum_i \sum_{\alpha_i} p_{\alpha_i} |\psi_{\alpha_i}\rangle\langle\psi_{\alpha_i}|, \quad \text{(S14)}$$

where $\alpha \subset \{0, 1, (\cdots), D-1\}$ with $|\alpha| \leq d$ (where $|\alpha|$ is the number of elements in $\alpha$) denotes the set of dimensions in which the decomposition element is entangled. Without loss of generality we consider the case $|\alpha| = d$, as every density matrix that can be decomposed into Schmidt rank $d' < d$ states is strictly contained in this definition (for some $\lambda_k = 0$). We write (S13) as

$$g_{kl}(\rho) = 4\frac{\sum_{\alpha_i:(k\in\alpha_i)\wedge(l\in\alpha_i)} \sum_i p_{\alpha_i} \lambda_k^{\alpha_i} \lambda_l^{\alpha_i}}{\sum_{\alpha_i:(k\in\alpha_i)\vee(l\in\alpha_i)} \sum_i p_{\alpha_i} ((\lambda_k^{\alpha_i})^2 + (\lambda_l^{\alpha_i})^2)} + 1. \quad \text{(S15)}$$

Here, $\lambda_k$ can be chosen to be real (as it is standard for the Schmidt decomposition), as any complex phases would only decrease the value of $Re[\lambda_k \lambda_l]$. Using the abbreviation $\tilde{\lambda}_k^{\alpha_i} := \sqrt{p_{\alpha_i}} \lambda_k^{\alpha_i}$ we can take the partial derivatives with respect to $\tilde{\lambda}_k^{\alpha_i}$ to find general conditions for all extremal points:

$$\frac{\partial}{\partial \tilde{\lambda}_k^{\alpha_i}} C(\rho) = \sum_{l\in\alpha} \frac{4\tilde{\lambda}_l^{\alpha_i}}{N_{kl}} - 2\tilde{\lambda}_k^{\alpha_i} \sum_l \frac{g_{kl}-1}{N_{kl}} = 0. \quad \text{(S16)}$$

Thus

$$\sum_l \frac{g_{kl}-1}{N_{kl}} = \sum_{l\in\alpha} \frac{4\tilde{\lambda}_l^{\alpha_i}}{N_{kl} 2\tilde{\lambda}_k^{\alpha_i}}. \quad \text{(S17)}$$

Since the left hand side is symmetric in $k$ and $l$ we can conclude that $\tilde{\lambda}_l^{\alpha_i} = \tilde{\lambda}_k^{\alpha_i}$. It then directly follows that

$$\sum_l \frac{g_{kl}-1}{N_{kl}} = \sum_{l\in\alpha} \frac{4}{2N_{kl}}. \quad \text{(S18)}$$

And since only the right hand side depends on $\alpha$, every partial sum of $\frac{1}{N_{kl}}$ is equal for $k, l \in \cup \alpha_i$. This implies that all $N_{kl}$ that appear in the union of all $\alpha_i$ will be equal to some $N$. Thus we can write

$$C(\rho_{max}) \leq \sum_{k,l\in\cup\alpha_i} \frac{f_{kl}}{N} + \sum_{k\in\cup\alpha_i, l\notin\cup\alpha_i} \frac{N_{kl}}{N_{kl}}, \quad \text{(S19)}$$

where $N = \frac{2}{|\cup\alpha_i|}$ and we know from from eq(S12) that $\sum_{k<l} \sum_{l=1}^{x-1} f_{kl} \leq 2d + (x-3)$ for all $x \geq d$. We thus get an upper bound

$$C(\rho) \leq Dd + \frac{D}{2}(D-3). \quad \text{(S20)}$$

Now we know that all extremal points (of course including all maxima) of the function are upper bounded by this value. At the same time we know that on every boundary (given by a partial set of $|\psi_{\alpha_i}\rangle$) all extremal points are upper bounded by an even lower value. Together this implies that indeed we have found the global maximum.

## Maximizing quantum state

It is worth pointing out that in general for every $d < D$ this bound is tight, i.e. there exists a mixed state which saturates this bound

$$\rho_d = \frac{1}{\binom{D}{d}} \sum_{\alpha \subset \{0,1,\cdots,D-1\}} |\phi_\alpha^d\rangle\langle\phi_\alpha^d|, \tag{S21}$$

where $|\phi_\alpha^d\rangle := \frac{1}{\sqrt{d}} \sum_{k \in \alpha} (|kk\rangle)$ and $|\alpha| = d$.

When we compare the maximizing state with the expected state in our experiment (non-maximally entangled pure state), we see that we most likely will underestimate the Schmidt number for photons created in down-conversion, thus might have an even higher number of entangled dimensions.

## EFFECT OF DEVIATION FROM PERFECT CORRELATION

For the derivation of the bounds (S21), we assumed perfect correlations between all involved modes. This is physically motivated: The first quantum number corresponds to the angular momentum of light, which is known to be conserved in the down-conversion process [3][4][5]. The second quantum number corresponds is the radial momentum, for which the limit of perfect-correlation (so-called "quasi-schmidt-modes") can be realized experimentally with very good approximation [6][7], as it can be seen in our experimental data in Fig. 3b.

Deviations from this assumption have been analysed numerically. We have analysed how non-perfect correlated states or non-orthogonal projections influence our quantity (S9). We have found that for small deviations such as observed in our experiment, the quantity (S9) can only decrease thus the resulting bounds still hold. Therefore the application of our entanglement dimensionality criterion is perfectly justified.

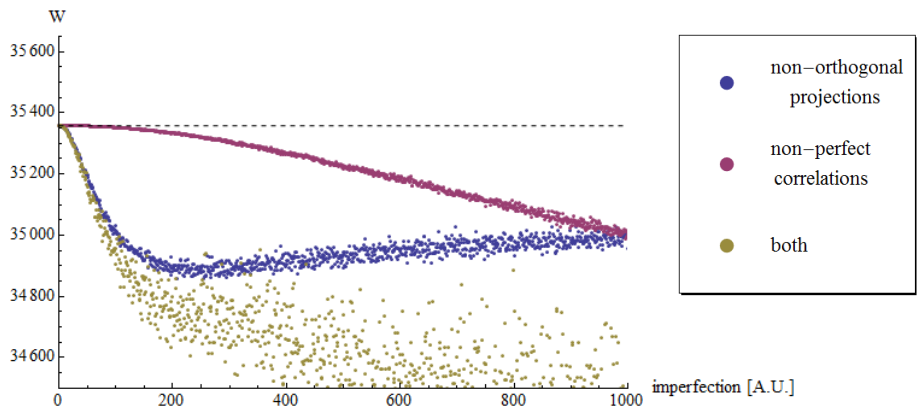

*FIGURE S1*: A non-maximally entangled state with D=186 dimensions is considered, similar to the state we expect from our experiment. It is considered how non-orthogonal projections and other measurement induced errors (blue), non-perfect correlations (red) and both effects simultaneously (yellow) influence the value of the quantity W ( or C($\rho$) S9). For each type, 1000 cases are calculated. The imperfections are introduced randomly, in each step the introduced imperfections increase. The dashed black line shows the value calculated without imperfections. It can be seen that any introduction of non-perfect correlations or non-orthogonal projections only decreases W. These results guarantee that the experimentally observed deviation from the perfectly correlated state can not artificially increase the observed dimensionality (in contrast, it decreases the quantity W, thus the observed Schmidt number).

## OBSERVED VISIBILITIES

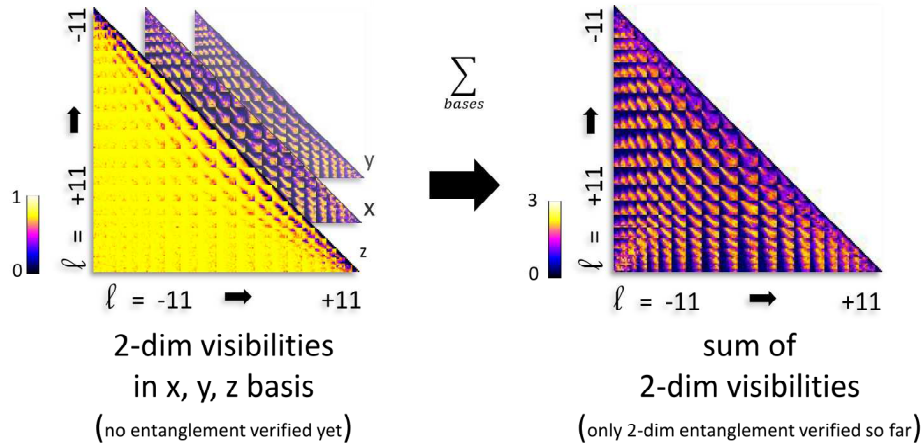

*FIGURE S2*: The left picture shows the visibilities of all the 2-dimensional subsets in all the three bases. The z-visibility is usually bigger, because non-maximal entanglement (Fig. 3a) reduces visibility in the x- and y-basis only. When we sum up the three visibilities (right picture), we can see that some subsets are more entangled than others. Every subspace with a value bigger than 1 is 2-dimensionally entangled; this is true for most of the 17.000 subspaces. Modes with similar count-rates have high visibilites, whereas modes with a very different count-rates have very low visibilies in the x- and y- basis. To reveal information about the global, high-dimensional entanglement, we have to sum up all visibilities of all subsets and calculate quantity W in eq.(1).

## CONSIDERED MODES VS. ENTANGLEMENT DIMENSIONALITY

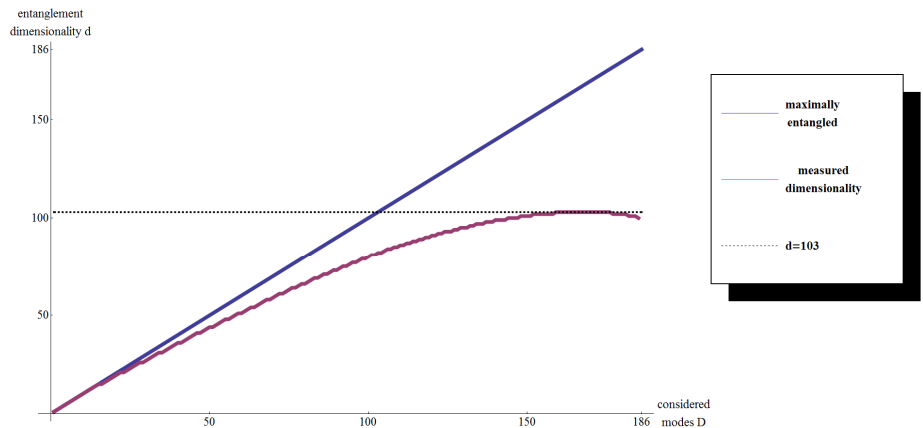

*FIGURE S3*: The entanglement dimensionality d depends on the number of modes considered for $<W>$, as described in inequality (2) in the main text. We measured all two-dimensional subspace correlations in x-, y-, and z-basis for 186 different modes. If the state would be maximally entangled, one could extract as much entanglement dimensionality as one considers modes (blue curve). However, as it can be seen in figure 4a (main text), there are modes which contribute stronger to $<W>$ (yellow region), and some modes that contribute less to the witness (for instance the Gaussian mode $LG_{0,0}$). One can remove certain modes (i.e. not considering all two-dimensional subspace measurement that contains these specific modes). We find a maximal detectable entanglement dimensionality d, if we only consider the 167 strongest contributing modes (red curve). This leads to 103-dimensional entanglement (black dashed line). We also observe a set of 15 modes which are entangled in 15 dimensions, which this could be significant in special protocols in quantum communication.

**Example:** To explain this behaviour more, we give a simple example. Considering the state

$$|\psi\rangle = N\Big(0.5|0,0\rangle + 0.07|1,-1\rangle + 0.01|2,-2\rangle + 0.01|3,-3\rangle\Big),$$

where $N$ is the normalization, we can calculate the subspace visibilities. The sum of the x,y,z visibilities in every subspace (SV=$V_x$+$V_y$+$V_z$) gives:

$SV^{0,1}$=1.55, $SV^{0,2}$=1.08, $SV^{0,3}$=1.08, $SV^{1,2}$=1.56, $SV^{1,3}$=1.56, $SV^{2,3}$=3

Calculating the sum of all visibilities (according to equation (1) in the main text), gives: W=9.723. Comparing with our inequality from (2) in the main text, using D=4, we can confirm a 2-dimensional entanglement (Bounds are 6, 10 and 14 for 2-dim, 3-dim and 4-dim entanglement, respectivly).

If we only consider mode 1, 2 and 3 (and don't consider mode 0), we get W=6.12, which confirms a 3-dimensional entanglement (Bounds are 3 and 6 for 2-dim and 3-dim entanglement, respectivly). This example shows how considering a smaller number of subsets can verify a higher-dimensional entanglement with our method.

## COMPUTER-GENERATED HOLOGRAMS

The holograms on the SLMs are calculated using a plane-wave approximation. This could lead to non-orthogonal projective measurements. However, for our system this effect can only reduce the visibilities and thus reduce the observed dimensionality. We restrict ourselves to two-dimensional subspaces, as this leads to simpler holograms on the SLMs and increases the mode transformation accuracy due to the finiteness of the pixels. Furthermore this method allows us to treat non-maximally entangled state (Figure 3a) directly; therefore we do not need to perform any entanglement concentration.

## STATISTICAL UNCERTAINTY

The detected photon numbers are assumed to be Poisson distributed, which leads to asymmetric distribution especially for low count rates. Analytical treatment of error propagation for such a big number of measurements was not feasible. Therefore, all confidence intervals have been calculated using Monte-Carlo simulations. The statistical uncertainty is very small because it depends on 200.000 measurements with a relatively small uncertainty.

### Acknowledgements

MH would like to acknowledge productive discussions of the entanglement detection criterion with Ariel Bendersky, Stephen Brierley, Jonathan Bohr-Brask, Daniel Cavalcanti, Ottfried Gühne, Karen Hovhannisyan, Claude Klöckl, Milan Mosonyi, Marcin Pawlowski, Martin Plesch, Paul Skrzypczyk and Andreas Winter.

---